\documentclass[referee,a4paper,12pt,traditabstract]{jswsc} 

\usepackage{graphicx}
\usepackage{txfonts}
\usepackage{subfigure}
\usepackage{epstopdf}
\usepackage[displaymath,mathlines]{lineno}
\usepackage[authoryear,round]{natbib}
\usepackage[backref]{hyperref}
\usepackage{url}
\usepackage{multirow}
\usepackage{gensymb}

\hypersetup{colorlinks=true,citecolor=cyan,urlcolor=cyan,linkcolor=blue}

\begin{document}


   \title{Variability in Footpoint Mapping of Bursty Bulk Flows Using Tsyganenko Models:}

   \subtitle{Impact on Swarm Conjunctions}
   
   \titlerunning{Footpoint Mapping of Bursty Bulk Flows}

   \authorrunning{V. Lanabere et al.}

\author{V. Lanabere
    \inst{1}\fnmsep\thanks{Corresponding author: \url{vanina.lanabere@irfu.se}}
    \and
    A. P. Dimmock\inst{1}
    \and
    L. Richard\inst{1}
    \and
    S. Buchert\inst{1}
    \and
    Y. V. Khotyaintsev\inst{1}
    \and
    O. Marghitu\inst{2}
 }

\institute{Swedish Institute of Space Physics, Institutet för rymdfysik Box 537, 751 21 Uppsala, Sweden
    \and
    Institute of Space Science, Bucharest, Măgurele 077125, Rumania
}


 
  \abstract{Magnetospheric-ionospheric coupling studies often rely on multi-spacecraft conjunctions, which require accurate magnetic field mapping tools. For example, linking measurements from the magnetotail with those in the ionosphere involves determining when the orbital magnetic footpoint of THEMIS or MMS intersects with the footpoint of Swarm. The Tsyganenko models are commonly used for tracing magnetic field lines. In this study, we aim to analyze how the footpoint locations are impacted by the input parameters of these models, including solar wind conditions, geomagnetic activity, and the location in the magnetotail. A dataset of 2394 bursty bulk flows (BBFs) detected by MMS was mapped to Earth's ionosphere with six different Tsyganenko models. Approximately 90\% of the ionospheric footpoints are concentrated within $70^\circ \pm 5^\circ$ magnetic latitude (MLAT) and $\pm 3$ hours of magnetic local time (MLT) around midnight, with a pronounced peak in the pre-midnight sector. The MLT position showed a difference of approximately $\pm 1$ hour MLT across the models. Footpoint locations were linked to the dawn-dusk position of the BBFs, with differences between models associated with variations in the interplanetary magnetic field clock angle. The MLAT values exhibited similar differences of approximately $\pm 4^\circ$ around the mean value, with a systematic shift toward lower latitudes in the T89 model. This position is also influenced by the input parameters of the model representing the dynamics of Earth's magnetosphere, where stronger magnetospheric activity typically corresponds to lower latitudes. The uncertainty on the BBF footpoint location impacts the number of conjunctions with Swarm. Generally, Swarm B exhibited more conjunctions than Swarm A or C in the Northern Hemisphere. However, when considering only Swarm-BBF conjunctions where the distance between footpoints computed with T89 and TA15n is smaller than the size of the BBF footprint, the number of conjunctions is reduced to less than half of the total.}


   \maketitle

\section{Introduction}
Substorms play a crucial role in the dynamics of the Earth's magnetotail, significantly influencing the transport of mass, energy, and magnetic flux. One of the key processes during substorms is the generation of transient high-speed plasma flows, known as bursty bulk flows \cite[BBFs,][]{Angelopoulos_1994}. These BBFs eventually couple to the high-latitude ionosphere by forming field-aligned currents (FACs) systems closed by an electrojet current in the ionosphere. This coupling has notable ionospheric and ground magnetic manifestations, such as auroral streamers and magnetic field disturbances ($dB/dt$) spikes \cite[and references therein]{Juusola_2009}.

To study the ionospheric and ground manifestation of magnetotail drivers, it is essential to utilize multi-point observations \cite[e.g.][]{Wei_2021, Aryan_2022, Ngwira_2025}. For example, linking measurements from the magnetotail to the ionosphere requires determining the times at which the orbital magnetic footpoints of different spacecraft intersect. In this context, accurately mapping spacecraft positions throughout the magnetosphere to specific locations in the ionosphere is crucial. The most commonly used external magnetic field models for connecting magnetospheric dynamic processes to their ionospheric signatures are the Tsyganenko models \cite[][hereafter T87]{Tsyganenko_1987}, \cite[][hereafter T89]{Tsyganenko_1989}, \cite[][hereafter T96]{Tsyganenko_1996}, \cite[][hereafter T01]{Tsyganenko_2002a, Tsyganenko_2002b}, and \cite[][hereafter T04]{Tsyganenko_2005}, and the more recent TA15 model \cite[][hereafter TA15]{Tsyganenko_2015} which has a more sophisticated representation of the magnetospheric system and uses a larger dataset of spacecraft measurements. 

Although model accuracy (i.e., how closely the derived footpoint matches its true value) is not the focus of this study, previous research has highlighted limitations in these models. For instance, \cite{Pulkkinen_1996} analyzed the accuracy of the T89 model, finding that the mapping errors were largest on the dayside, while on the nightside they ranged from $0.5\degree$ to $1.8\degree$ in latitude ($63\degree–72\degree$), increasing with geomagnetic activity. Similarly, \cite{Shevchenko_2010} reported errors of $\sim1\degree$ when evaluating the accuracy of the T96 model. Thus, the T96 model performs a better mapping than T89, but in both cases the mapping error increases with magnetic activity. These discrepancies have been observed in case studies. For instance, \cite{Opgenoorth_1994} found that a localized FAC was observed a few degrees north compared to the location obtained with the T89 model. Similarly, \cite{Wei_2021} noted differences in the ground track mapping when using three different Tsyganenko models (T87, T89, and T96) while investigating the characteristics and responses of the magnetosphere-ionosphere-ground system during a case study.

This study utilizes a large database of BBF events observed by the Magnetospheric Multiscale (MMS) spacecraft \citep{Burch_2016} to statistically compare the mapped footpoint locations among six different Tsyganenko models. We refer to the footpoint as a unique point representing the BBF and do not consider the spatial size of the BBF mapped into the ionosphere (footprint). In this work, we do not address the error in the location of the footpoint, since this implies establishing a ground truth for the correct footpoint for each event, which is beyond the scope of this analysis. Instead, we analyze how the footpoint locations are impacted by the input parameters of these models, including solar wind conditions, geomagnetic activity, the location of the BBF in the magnetotail, and the tilt angle. We report how the number of conjunctions between BBFs and Swarm spacecraft changes when the conjunction definition is adjusted. Specifically, we examine how increasing the time interval between the detection of a BBF and the Swarm satellites' path through/nearby the BBF footpoint, as well as varying the footpoint magnetic latitude error, affects the occurrence and quality of conjunctions. Although we did not conduct a study on the error in the location, we hope that the quality indicator will help researchers determine when to be more cautious when studying conjunctions.

\section{Data and models}

\subsection{The BBF database}

We used the BBF database developed by \cite{Richard_2022db} which consists of the initial and end time of 2394 BBFs detected in the magnetotail by MMS 1 during the five magnetotail seasons between 2017-2021. A full description of the selection criteria can be found in \cite{Richard_2022}.

Since our goal is to find the ionospheric footpoint of the BBFs, we restricted our database to Earthward-moving BBFs, which constitute 89\% of the total database. The mean and median duration of these Earthward BBFs are 3.60 minutes and 2.63 minutes, respectively, with approximately 80\% of the events lasting less than 5 minutes. After excluding an outlier with a duration of $\approx 3~$hours, the BBF dataset consists of 2135 events, with the longest event lasting 35~minutes. 

\begin{figure}
    \centering
    \includegraphics[width=\linewidth]{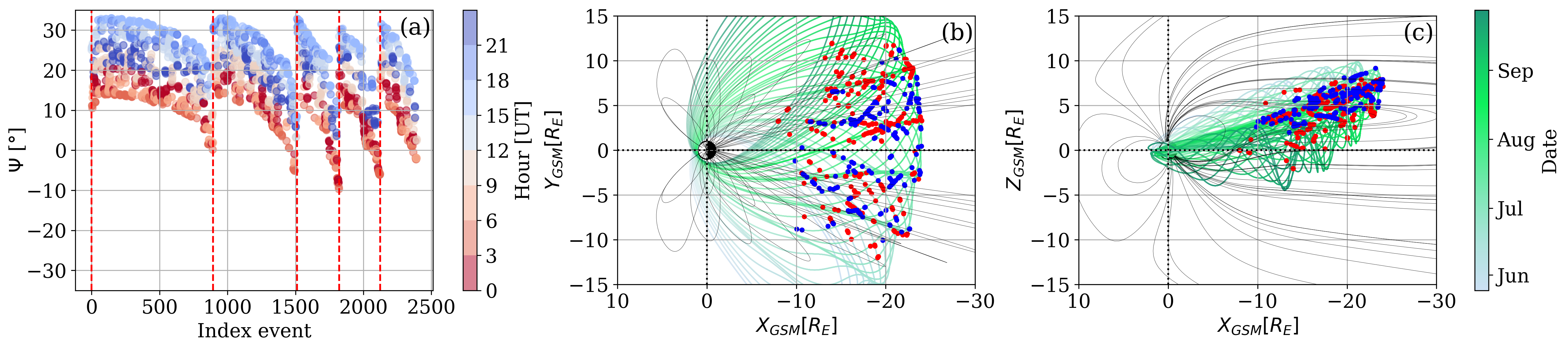}
    \caption{Observational bias of Earthward BBFs events related to the MMS tail/summer season. (a) Dipole tilt angle at the moment of the 2135 BBF event chronologically ordered. The dashed red lines correspond to the dates of the first BBF detected during each MMS tail summer season: 2017-05-16, 2018-05-30, 2019-06-18, 2020-06-27, and 2021-06-26. (b-c) BBF position in GSM coordinates for the BBF event detected during the second MMS tail season (red and blue dots) and when $20^\circ<\Psi<30^\circ$ (blue dots). The background magnetic field lines are produced with T89 for $\Psi=25^\circ$, Kp=1, $V_\mathrm{x,GSE}=-400~\mathrm{km/s}$.}
    \label{fig_database_bias}
\end{figure}

Due to the MMS orbit, there is an observational bias in the detected BBFs, as illustrated in Figure \ref{fig_database_bias}. The first panel shows that the dipole tilt angle ($\Psi$) at the moment of the BBF detection is predominantly positive (mean $\Psi=18^\circ$), peaking in June and gradually decreasing toward the end of the MMS tail season. Superimposed on this annual variation is the daily variation, with the largest $\Psi$ values occurring between 15:00 and 18:00~UT (Figure \ref{fig_database_bias}a). 

Additionally, Figure \ref{fig_database_bias}b-c show the MMS orbit for the second MMS season (2018-05-25 to 2018-09-28) and the BBFs detected during the same period (blue dots correspond to BBF detected when $20^\circ<\Psi<30^\circ$). In the $Y_\mathrm{GSM}$ coordinate, the MMS apogee begins the magnetotail season at $Y_\mathrm{GSM}<0$, shifting to $Y_\mathrm{GSM}>0$ by the end of the season. Similarly, the apogee in $Z_\mathrm{GSM}$ starts positive and transitions to negative as the season progresses, in line with the tilt angle.


\subsection{Solar wind and ground magnetic conditions}

For determining the footpoints, it is necessary to incorporate data on solar wind and geomagnetic conditions. We used 1-minute resolution solar wind parameters, including solar wind dynamic pressure ($P_{dyn}$), interplanetary magnetic field (IMF), and solar wind velocity in GSE coordinates ($V_{i}$), as extracted from NASA/GSFC's OMNI data set through OMNIWeb. The geomagnetic Dst index was replaced by the higher resolution Sym-h index, and the Kp index was replaced by the higher resolution Hp30 index \citep {Yamazaki_2022}. Additionally, the solar wind driving parameters $G$ and $W$ \citep{Qin_2007} have been used as input parameters for T01 and T04 respectively, as well as the solar wind coupling functions B-index \citep{Boynton_2011} and N-index \citep{Newell_2007} for the TA15b and TA15n models, respectively. 

The solar wind conditions for each event were determined by calculating the mean values within a 30~minutes window centered on each BBF event. The 30-minute window was chosen to capture the mean solar wind conditions that reflect the mean magnetospheric configuration in the interval from the solar wind driver to the ionospheric signature. A possible propagation delay from the bowshock to the BBF location in the magnetotail was neglected. Events lacking solar wind data within this time window were excluded from the analysis, as these data are necessary for running the mapping models. Consequently, the BBFs database was reduced to 2079 BBF events. 

The geomagnetic conditions during the BBFs events are characterized by a mean Sym-h index of $-10~\mathrm{nT}$ and a median of $-8~\mathrm{nT}$. Notably, 72\% of the events have Sym-h values between $-30~\mathrm{nT}$ and $0~\mathrm{nT}$, which is above the threshold typically associated with weak geomagnetic activity. This is consistent with the distribution of the Hp30 index, where 78\% of events have $\mathrm{Hp30}<3$.


\subsection{Mapping models}

The footpoints for each of the 2079 BBF events were computed using the Geopack tracing routine for the Tsyganenko models T89, T96, T01, and T04, available through the Geopack Python library\footnote{\url{https://github.com/tsssss/geopack}}, and TA15b, TA15n, available through the IDL Geopack DLM\footnote{\url{https://korthhaus.com/idl-software/idl-geopack-dlm/}}. Each model requires the event date and solar wind velocity to calculate the dipole tilt angle. The input parameters and valid regions for each model are summarized in Table~\ref{tab:tsyganenko_input}.

\begin{table}
    \caption{Kp and Dst are geomagnetic indices; $P_{dyn}$ the dynamic pressure of the solar wind; $B_\mathrm{y,IMF}$ and $B_\mathrm{z,IMF}$ the y and z components of the IMF; G1 and G2 are the parameters related to solar wind and IMF conditions from the preceding 1-hour interval; W1 to W6 are the solar wind driving variables, quantifying the magnitudes of principal magnetospheric current systems; B-Index and N-index are solar wind coupling function, suggested by \cite{Boynton_2011} and \cite{Newell_2007} respectively. $r_\mathrm{GEO}$ is the geocentric distance}
    \centering
     \renewcommand{\arraystretch}{1.2} 
    \setlength{\tabcolsep}{10pt} 
    \begin{tabular*}{\textwidth}{l@{\extracolsep{\fill}}cc}
    \hline
    Model & Inputs & Valid region \\
    \hline
    T89  & Kp & $r_{\rm{GEO}} \leq 70 R_E$ \\
    T96 & Dst,$P_\mathrm{dyn}$ nPa,
    $B_\mathrm{y,IMF}$, $B_\mathrm{z,IMF}$,  & $r_{\rm{GEO}} \leq 60 R_E $ \\
    T01  & Dst, $P_\mathrm{dyn}$, $B_\mathrm{y,IMF}$, $B_\mathrm{z,IMF}$, $G_1$, $G_2$ & $X_{\rm{GSM}} \geq -15 R_E$ \\
    T04  & Dst, $P_{dyn}$, $B_\mathrm{y,IMF}$, $B_\mathrm{z,IMF}$, $W_1$ to $W_6$ & $X_{\mathrm{GSM}} \geq -15 R_E$ \\
    TA15b  & $P_{dyn}$, $B_\mathrm{y,IMF}$, $B_\mathrm{z,IMF}$, $B\mathrm{-Index}$ & $r_{\rm{GEO}} \leq 60 R_E $\\
    TA15n  & $P_{dyn}$, $B_\mathrm{y,IMF}$, $B_\mathrm{z,IMF}$, $N\mathrm{-Index}$ & $r_{\rm{GEO}} \leq 60 R_E $ \\
    \hline
    \end{tabular*}
    \label{tab:tsyganenko_input}
\end{table}

The valid ranges for each input parameter are provided in the respective documentation for each model. However, we emphasize that the T01 and T04 models are only valid sunward of $X_{\rm{GSM}}=-15~R_E$. Consequently, only 12.5\% of the 2079 BBF events satisfy this criterion and can be mapped to the Earth's surface. Additionally, after visually inspecting the traced field lines, two criteria were applied to exclude events that could not be accurately modeled (see examples in the Appendix \ref{appendix_B}). The behavior of the models for such excluded events is beyond the scope of this work. The two criteria that were used to retain valid events are as follows:

\begin{enumerate}
    \item The footpoint in the Northern and Southern Hemispheres reaches the Earth's surface.
    \item The magnetic field line goes out and returns only once, making a single loop.
\end{enumerate}

The database of 2079 BBF events, including the Tsyganenko model inputs and the resulting magnetic footpoints in both the Northern and Southern Hemispheres for the T89, T96, T01, T04, TA15b, and TA15n models, is available in \cite{Lanabere_2024}. The footpoints at the Earth’s ionosphere are given in both GSM coordinates and the altitude-adjusted corrected geomagnetic (AACGM) coordinate system.

\section{BBF footpoints from different Tsyganenko models}

\subsection{Statistical footpoint positions for each Tsyganenko model}

\begin{figure}
    \centering
    \includegraphics[width=\linewidth]{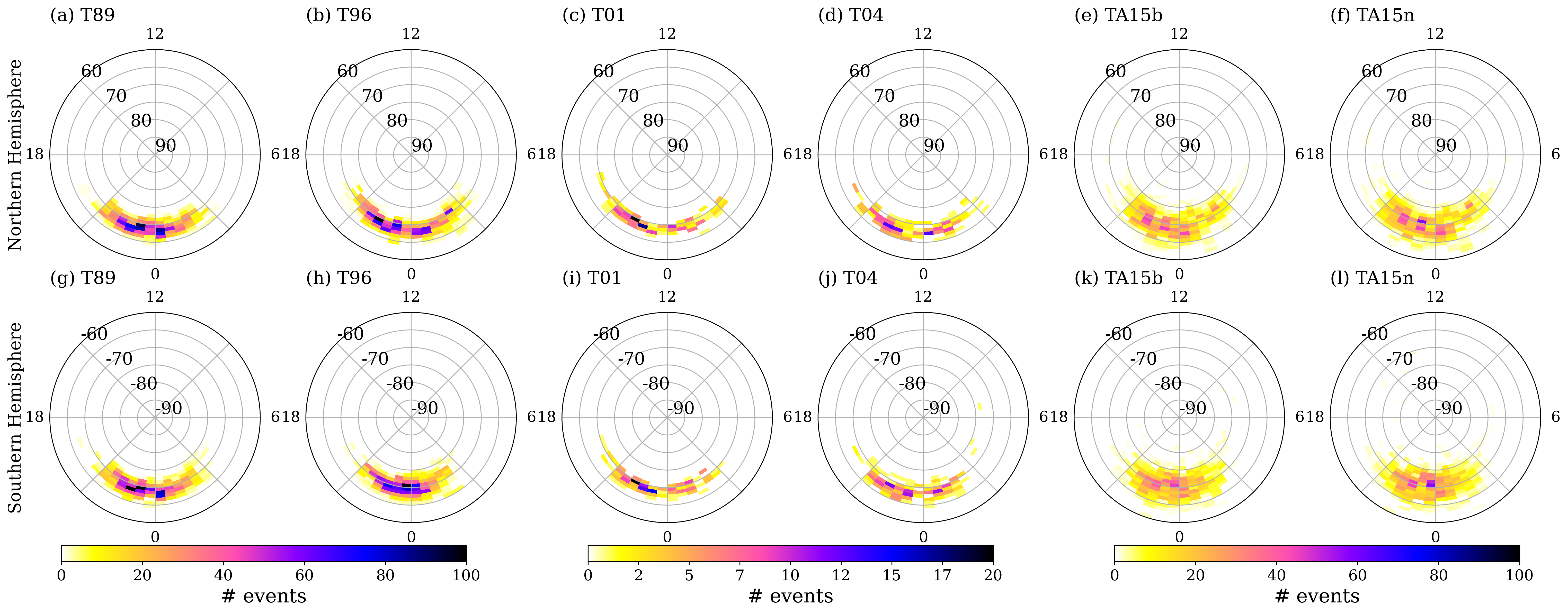}
    \caption{BBF ionospheric footpoint histograms in MLAT and MLT (AACGM coordinates) for different Tsyganenko models in the Northern and Southern Hemispheres. Bin size: $1^\circ$ x 0.5 hour.}
    \label{fig_hist_foorprint}
\end{figure}

We analyzed the footpoint of the BBF events in both the Northern Hemisphere and Southern Hemisphere (NH and SH respectively) for each Tsyganenko model. The footpoint histogram for each Tsyganenko model is shown in Figure \ref{fig_hist_foorprint}. Statistically, the footpoints appear rather similar across all the mapping models, suggesting that despite the inherent differences in how these models represent Earth's magnetosphere, they statistically converge on a similar spatial distribution of BBFs.

The magnetic latitude (MLAT) footpoints of BBFs are generally concentrated around $70^\circ \pm 5^\circ$ in both hemispheres. The wider MLAT spread observed in the TA15 model may be attributed to its capability to represent the magnetosphere under a wider variety of solar wind conditions. In terms of magnetic local time (MLT), BBFs predominantly have their ionospheric footpoints around midnight $\pm 3$ hours, with a clear peak in the pre-midnight sector presumably associated with substorms. 

While the statistical distribution of BBF footpoints is similar across the Tsyganenko models, significant variations can occur for specific events. These differences are crucial when searching for conjunctions. To better understand these differences, we performed a comparative analysis of the footpoint locations produced by each Tsyganenko model, using the TA15n model as the reference. TA15n was chosen as the benchmark because it maps a large number of BBFs, due to its broader inclusion criteria, and is expected to be more accurate due to its more sophisticated representation and larger dataset.

\subsection{Comparison of BBF footpoints: TA15 vs. previous models}
\begin{figure}
    \centering
    \includegraphics[width=\linewidth]{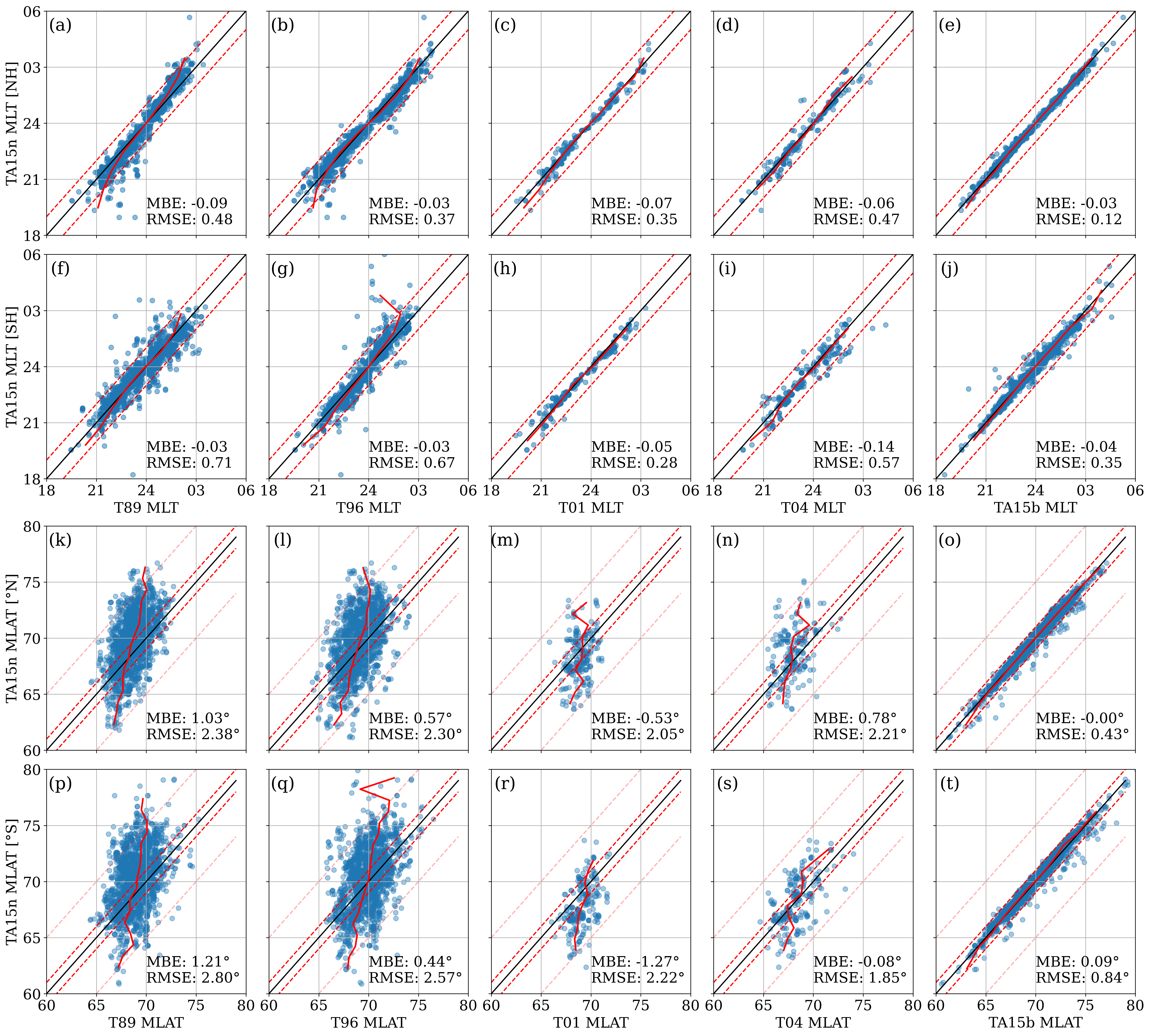}
    \caption{Comparison of footpoint position in AACGM coordinates for Ta15n versus the T89, T96, T01, T04, and TA15b. (a-j) Magnetic local time (MLT). (k-t) Magnetic Latitude (MLAT). The black diagonal line represents the ideal scenario where both models agree perfectly. The red dashed lines indicate a deviation of $\pm 1$ (hour for the MLT plots and degree for MLAT plots). The pink dashed line marks a $5^\circ$ difference in latitude, corresponding to the approximate distance of 1 hour of MLT at $70^\circ$ latitude. The solid red line represents the mean value. For each plot the mean bias error (MBE) and the root mean square error (RMSE) are indicated.}
    \label{fig_T89-T**}
\end{figure}

The MLT footpoint positions for each hemisphere are shown in Figure \ref{fig_T89-T**}(a–j), comparing the TA15n model with the T89, T96, T01, T04, and TA15b models. Additionally, the mean bias error (MBE) and the root mean square error (RMSE) are included to quantify the differences between the models.

The MLT positions between the models are quite similar, with differences typically within $\pm 1$ hour MLT. At $70^\circ$ latitude, this corresponds to a distance of approximately 571 km. However, as shown in Figure \ref{fig_T89-T**}a,f T89 tends to shift footpoints toward earlier MLT in the post-midnight sector and toward later MLT in the pre-midnight sector, relative to the TA15n footpoints in both hemispheres. A similar, though less pronounced, trend is observed between T96 and TA15n in the Southern Hemisphere, while the opposite trend is observed in the Northern Hemisphere.

The MLAT footpoint position is shown in Figure \ref{fig_T89-T**}(k-t) together with the MBE and RMSE. The differences between models in MLAT are of the same order of magnitude as those in MLT. To put this into perspective, approximately $5^\circ$ in latitude corresponds to the same distance as 1 hour of MLT at $70^\circ$, as indicated by the pink dashed line. In general, the TA15 models show a wider spread in MLAT than older models. The MBE between TA15n and T89 is $\mathrm{MBE}=1.03^\circ$ and $\mathrm{MBE}=1.21^\circ$ for the Northern and Southern Hemispheres, respectively, suggests that the MLAT positions calculated using TA15n are, on average, $1^\circ$ closer to the pole than those from T89. A similar poleward bias is observed when comparing TA15n to T96, though with smaller MBE values ($\mathrm{MBE}=0.57^\circ$ and $\mathrm{MBE}=0.44^\circ$ for the Northern and Southern Hemispheres). In contrast, the comparison between TA15n and T01 shows the opposite trend, with T01 footpoints located closer to the pole on average. No consistent latitudinal bias is observed between TA15n and T04. Due to the broader spread in MLAT for the TA15 model compared to older models, the RMSE values show that the MLAT footpoint location derived from TA15 and from (T89,T96, T01, T04) disagree by about $2^\circ$ of latitude on average, with the largest discrepancies observed when compared to T89.

\subsection{Relation between footpoint positions and input parameters}

\begin{figure}
    \centering
    \includegraphics[width=\linewidth]{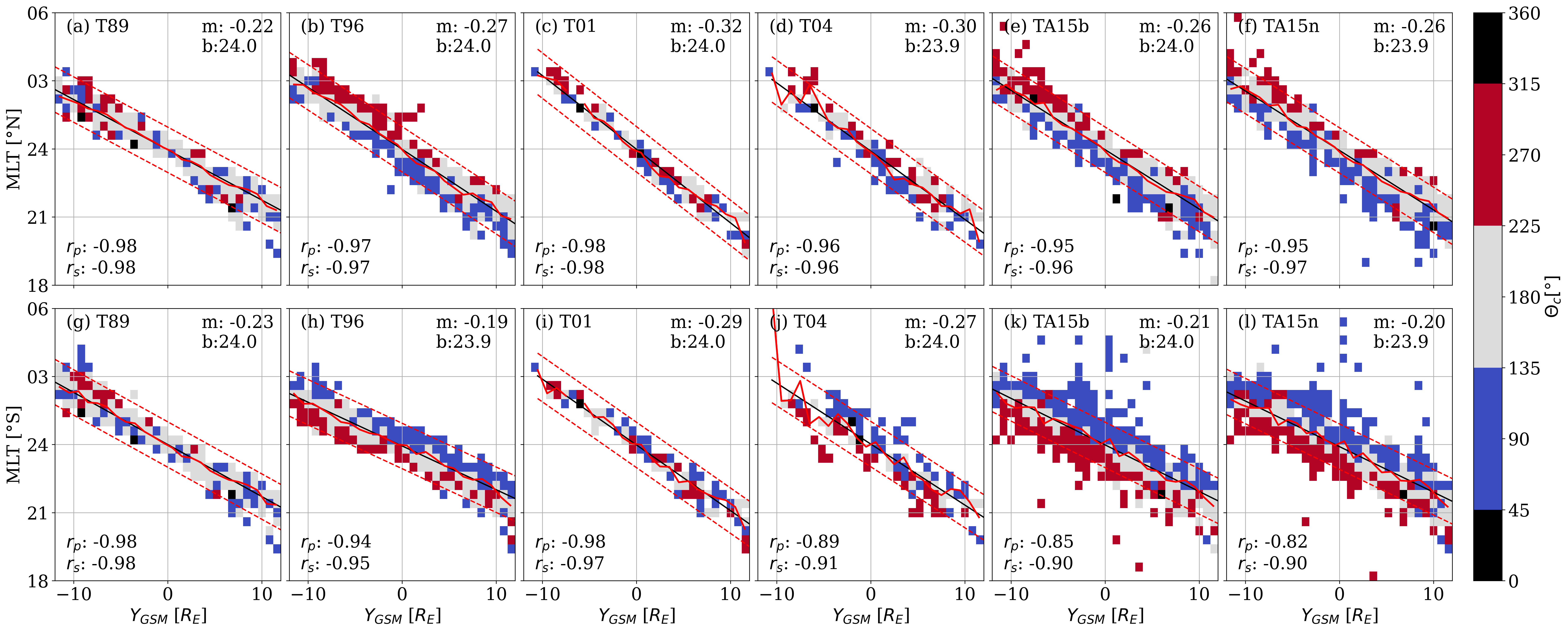}
    \caption{Relation between the BBF position ($Y_\mathrm{GSM}$) and MLT for each Tsyganenko model. The solid black line is the linear regression, while the dashed red line is the linear regression shifted $\pm 1$ hour. The solid red line indicates the mean value. Panels (a-d) correspond to the Northern Hemisphere, and panels (e-h) correspond to the Southern Hemisphere. The colorbar represents the IMF clock angle: Blue: IMF pointing towards dusk, Red: IMF pointing towards dawn, Gray: IMF pointing northward, and Black: IMF pointing southward.}
    \label{fig_mlt_ygsm}
\end{figure}

In terms of MLT, the strongest correlation was observed with the BBF position $Y_\mathrm{GSM}$ (see Figure \ref{fig_mlt_ygsm}), with Pearson and Spearman correlation coefficients around $-0.9$ in all cases. As expected, the linear coefficients indicate that on average pre-midnight events are associated with $Y_\mathrm{GSM}>0$, while post-midnight events correspond to $Y_\mathrm{GSM}<0$. Furthermore, the T96, T04, TA15b, and TA15n models that use $B_\mathrm{y,IMF}$, $B_\mathrm{z,IMF}$ as input parameters show a relation between MLT and the IMF clock angle ($\theta_c = \rm{atan}(B_y/B_z)$). For example, in the Northern Hemisphere, if $\theta_c\approx90^\circ$ ($B_{y,\mathrm{GSM}}>0$, pointing toward dusk), the MLT is earlier compared to $\theta_c\approx270^\circ$ ($B_{y,\mathrm{GSM}}<0$, pointing toward dawn). The opposite behavior is observed in the Southern Hemisphere. However, for the T01 model, this correlation is absent, which may be attributed to differences in the way models handle IMF components.

\begin{figure}
    \centering
    \includegraphics[width=\linewidth]{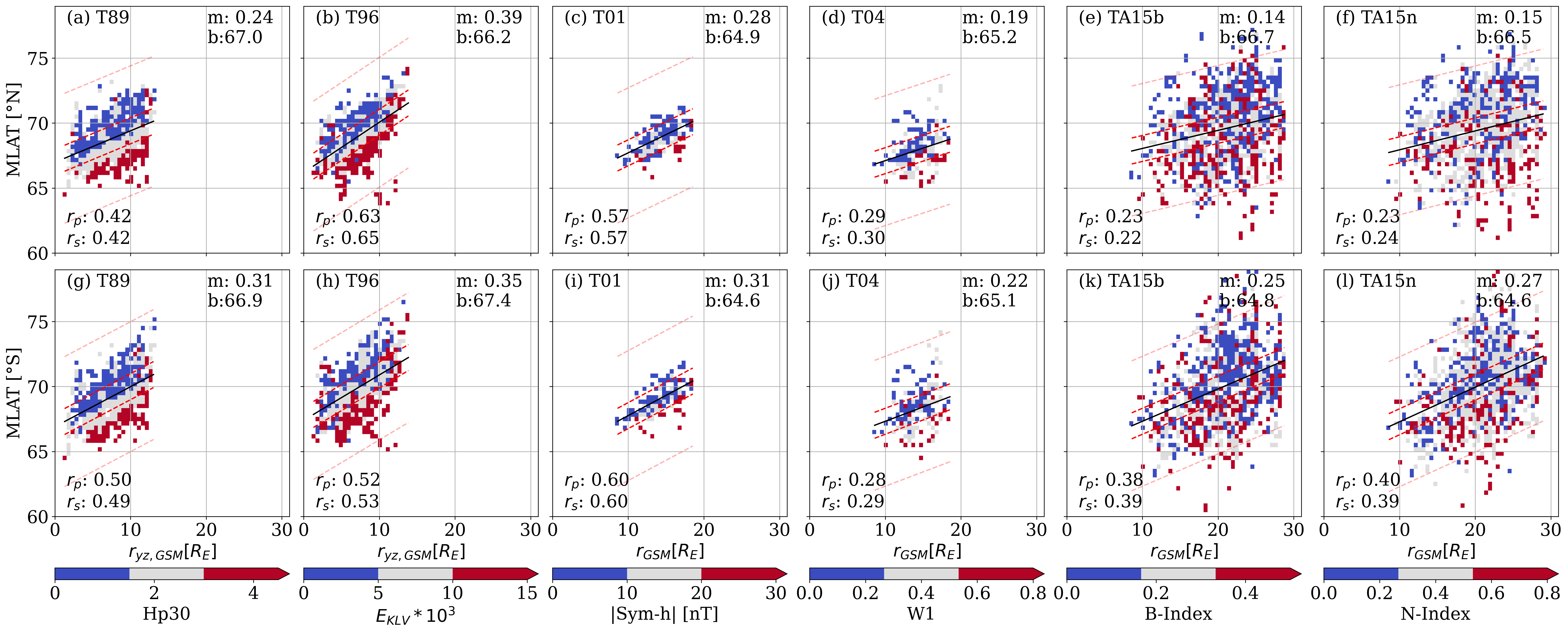}
    \caption{Relation between the BBF distance to Earth and MLAT for each Tsyganenko model. The color scale indicates the input variable or derived variable from the input parameters, with the largest correlation with MLAT. Panels (a-d) correspond to the Northern Hemisphere, and panels (e-h) correspond to the Southern Hemisphere. Each panel also displays the Pearson and Spearman correlation coefficients for MLAT vs. distance represented by the x-axis. The solid black line is the linear regression, while the dashed red and pink lines are the linear regression shifted $\pm 1^\circ$ and $\pm 5^\circ$ respectively.}
    \label{fig_mlat_pos}
\end{figure}

The relationship between MLAT and the input parameters of the models is more complex. As expected, cross-correlation analysis between the input variables and MLAT reveals a moderate correlation with the distance of the BBF from Earth, indicating that the MLAT shifts closer to the pole as the distance from Earth increases. Other parameters that influence MLAT vary depending on the specific model used. The Pearson and Spearman correlation coefficients between MLAT and the distance to Earth, $r_P$ and $r_S$, for each model and hemisphere are presented in Figure \ref{fig_mlat_pos}.

For the simplest model (T89), BBFs detected at similar distances from Earth exhibit footpoints that reach lower magnetic latitudes as the Hp30 values increase. A similar behavior is observed with T96 but with the coupling function $E_{KLV}= V^{4/3}B_T\mathrm{sin}^2(\theta_c/2)p^{1/6}$ proposed by \cite{Vasylundas_1982}. This coupling function relates all the solar wind input parameters and presents a larger correlation than the Sym-h index ($r_p=-0.18$ and $r_s=-0.08$). In the case of T01, the correlation with distance is also evident, but no clear relationship with input parameters was found, possibly due to the limited number of events. In contrast, the T04, TA15b, and TA15n models exhibit weaker correlation with Earth's distance. Furthermore, the TA15 models show no significant relationship with either the B-index or the N-index.

\subsection{Analysis of model differences based on input parameters}
\begin{figure}
    \centering
    \includegraphics[width=\linewidth]{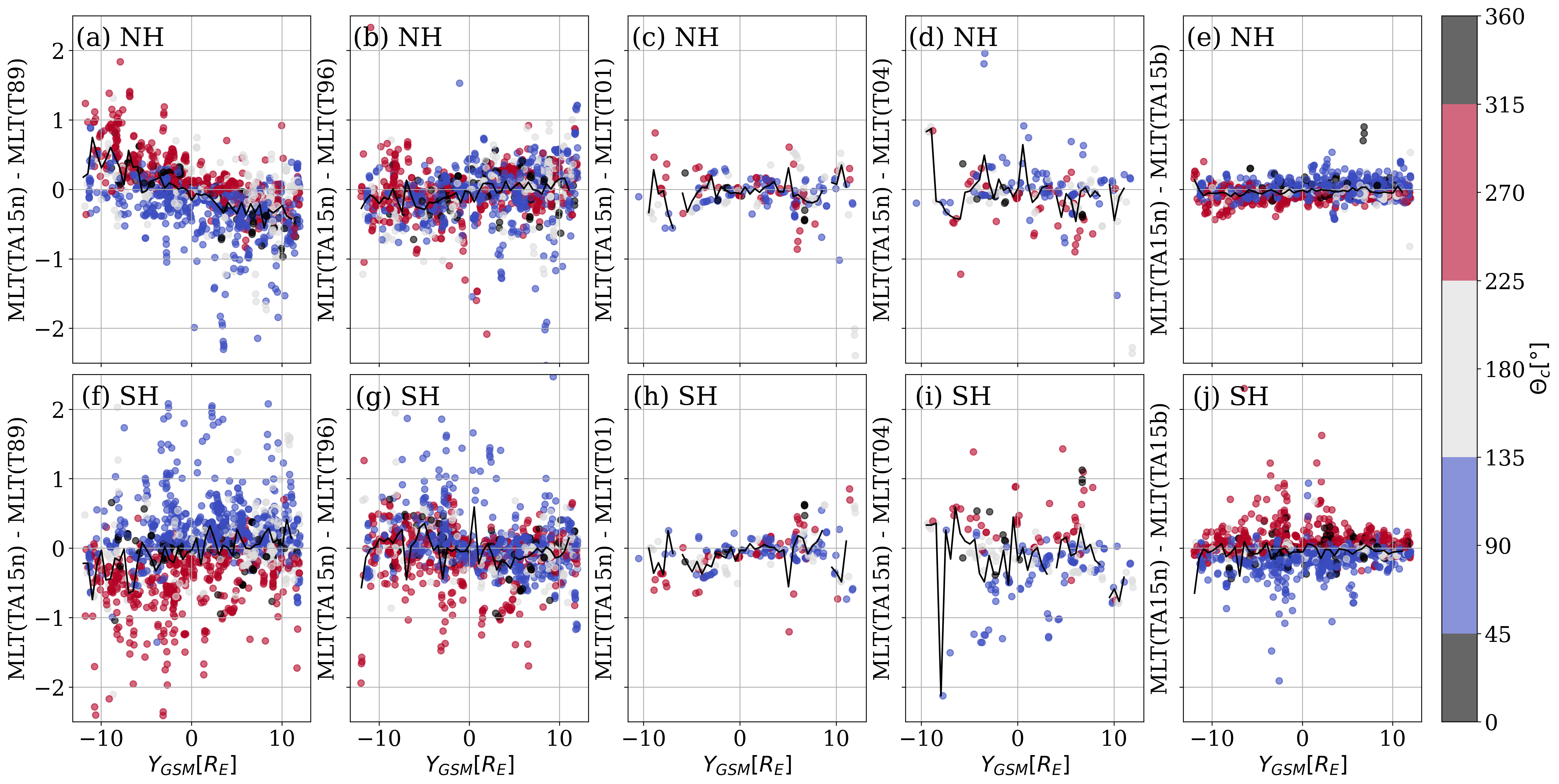}
    \caption{MLT difference between TA15n and (T89, T96, T01, T04, and TA15b) against the $Y_\mathrm{GSM}$ position of the BBF. The color scale is interplanetary magnetic field clock angle ($\theta_c$)  and the solid black line is the mean value.}
    \label{fig_MLT_error}
\end{figure}

The MLT difference between TA15n and (T89, T96, T01, T04, and TA15b) as a function of $Y_{GSM}$ and the clock angle $\theta_c$ is shown in Figure \ref{fig_MLT_error}. In the Northern Hemisphere, the T89 model tends to display footpoints closer to midnight compared to TA15n, while the opposite is observed in the Southern Hemisphere. This behavior is not clearly observed when comparing to the other models (T96, T01, T04, TA15b). The analysis of the dispersion of the clock angle around the mean error shows that when $\theta_c = 90^\circ$ (blue dots) the MLT mapped by T89 is later than that of TA15n, and earlier for $\theta_c = 270^\circ$ (red dots). An opposite trend is observed when comparing TA15n with TA15b, although the differences are smaller. A similar pattern appears in the comparison between TA15n and T04, though limited by low statistical coverage. In the Southern Hemisphere, the behavior is reversed to that observed in the Northern Hemisphere.

\begin{figure}
    \centering
    \includegraphics[width=\linewidth]{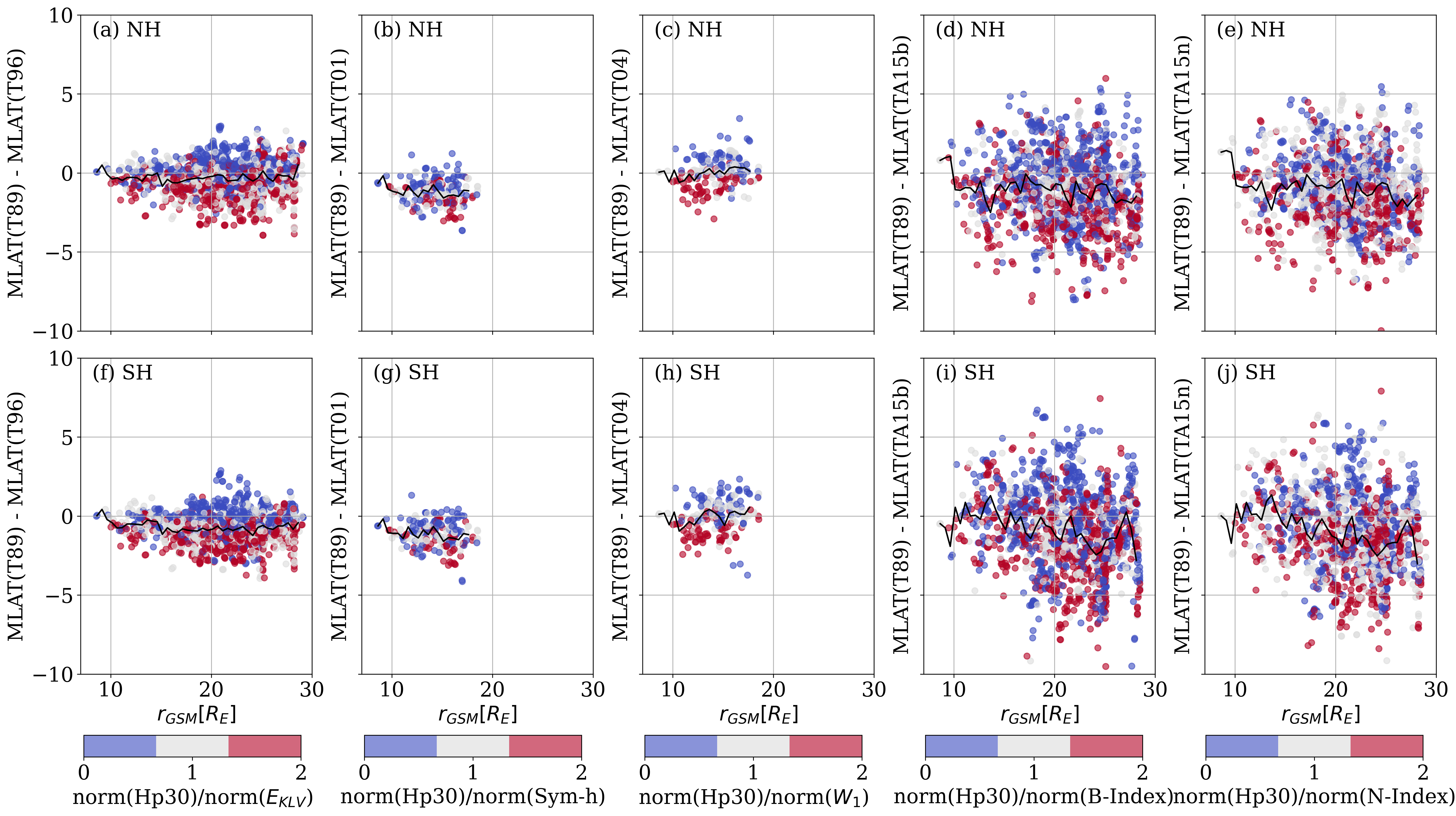}
    \caption{MLAT difference between T89 and (T96, T01, T04, TA15b, and TA15n) against the $r_\mathrm{GSM}$ position of the BBF. The color scale represents the ratio of the normalized Hp30 geomagnetic index (input parameter in T89) to the normalized input parameters with the highest correlation ($E_{KLV}$, Sym-h, $W_1$, B-index, and N-index for T96, T01, T04, TA15b, and TA15n respectively). The solid black line indicates the mean value. Normalization was performed using the median value.}
    \label{fig_MLAT_error}
\end{figure}

While previous results focused on comparisons with TA15n, the following analysis uses T89 as the reference model, since TA15n shows no clear relationship with Earth distance or magnetospheric activity proxies, unlike the older models, as shown in Figure \ref{fig_mlat_pos}. Accordingly, Figure \ref{fig_MLAT_error} presents the MLAT differences between T89 and (T96, T01 and T04, TA15b, and TA15n) as a function of the distance of the BBF from Earth and the relative magnitudes of the proxies for Earth’s magnetospheric dynamics. This ratio indicates how the input parameter for T89, as represented by Hp30, relates to $E_{KLV}$, Sym-h, $W_1$, B-index, and N-Index for the models T96, T01, T04, TA15b, and TA15n respectively, after normalizing by their median values. In the Northern Hemisphere, for BBFs located at the same $r_\mathrm{GSM}$, when Hp30 is greater than $E_{KLV}$ (red dots), T89 footpoints are at lower latitudes compared to T96 (Figure \ref{fig_MLAT_error}a). Conversely, when $E_{KLV}$ is greater than Hp30 (blue dots), T89 footpoints are closer to the pole than the ones computed with T96. The same behavior is observed in the Southern Hemisphere Figure \ref{fig_MLAT_error}f. A similar trend is observed when comparing T89 with T04, although it is less apparent with T01, possibly due to lower statistics. As mentioned earlier, no clear trend is observed in the comparisons involving the TA15 models.

\subsection{Footpoints proximity as a measure of mapping quality}

The difference between the footpoints, computed with the T89 and TA15n models (Figure \ref{fig_MLT_error}a,f and Figure \ref{fig_MLAT_error}a,f) shows that more than 80\% of the BBF footpoints in both hemispheres have an absolute MLT error less than $1~\mathrm{hour}$ and an absolute MLAT error less than $4^\circ$. At $70^\circ$ latitude, these correspond to an east-west extent of approximately $568~\mathrm{km}$ and a north-south extent of $444~\mathrm{km}$, respectively. These differences are significantly larger than the estimated mapped mean dimensions of dipolarization flux bundles (DFBs), which are typically associated with BBFs. According to \cite{Engebretson_2024}, a DFB with a mean radius of $0.9R_E$ \citep{Liu_2013b} mapped to the ionosphere has an east-west extent of approximately $180~\mathrm{km}$ ($0.3~\mathrm{MLT}$ at $70^\circ\mathrm{N}$) and a north-south extent of about $90~\mathrm{km}$ ($0.81^\circ$). This is consistent with our estimation of the BBF footprint size, as detailed in the Appendix \ref{appendix_A}. Assuming a dipole field model and a circular cross section, we calculated a BBF cross-section in the ionosphere of approximately $110~\mathrm{km}\pm30\mathrm{km}$.

Due to both statistical and physical considerations of the footpoint distances, we used two different criteria to assign a quality metric to the mapping. Based on the statistical difference between the footpoints calculated with the T89 and TA15n models, we define a maximum distance between footpoints of $\pm 1$ hour in MLT and $\pm 4^\circ$ in MLAT. Meanwhile, considering the physical size of the footprint, the maximum distance is 0.3~MLT and $0.81^\circ$. We introduce two footpoint quality metrics for each BBF mapped event: $Q_{fp}^{\rm{stat}}$, based on the statistical criterion, and $Q_{fp}^{\rm{phys}}$, based on the physical footprint size. In both cases, $Q_{fp}=1$ indicates that the footpoint distances between T89 and TA15n fall within the defined thresholds for both coordinates; $Q_{fp}=2$ indicates that only one coordinate is within threshold; and $Q_{fp}=3$ indicates that both coordinates fall outside the defined limits. In the first case, for the statistical criterion, the fraction of events with $Q_{fp}^{\rm{stat}}= \{1, 2, 3\}$ are 89.0\%, 9.6\% and 1.4\%, respectively, for the Northern Hemisphere, and 79.5\%, 18.7\%, and 1.8\%, respectively, for the southern hemisphere. In the second case, where the BBF footprint size is considered, the results for $Q_{fp}^{\rm{phys}}= \{1, 2, 3\}$ are $23.8\%$, $47.7\%$ and $28.5\%$ for the Northern Hemisphere, and $14.6\%$, $51.8\%$, and $33.6\%$ for the Southern Hemisphere, respectively.


\section{BBF and Swarm conjunctions}

In the previous section, we observed that the BBF footpoint locations predicted by different Tsyganenko models exhibit variations and that these discrepancies are influenced by geomagnetic conditions and the specific position where the BBF was detected. Consequently, defining conjunctions can be challenging due to these variations in model predictions. In this section, we want to implement the analysis we have discussed to investigate the impact on MMS-Swarm BBF conjunctions.

To address this, we will analyze how the number of detected conjunctions changes with different definitions of conjunction criteria. Specifically, we will investigate the impact of varying the MLAT error ($Err_\mathrm{MLAT}$) and the time interval between the detection of BBFs and the Swarm satellites' paths ($\Delta t$). The MLT error is set at $\pm 1~\mathrm{hour}$ because it corresponds to the typical dispersion of the BBF footpoints for different solar wind and magnetospheric conditions. This analysis aims to quantify how adjustments in these parameters influence the identification of conjunctions.

\begin{figure}
    \centering
    \includegraphics[width=\linewidth]{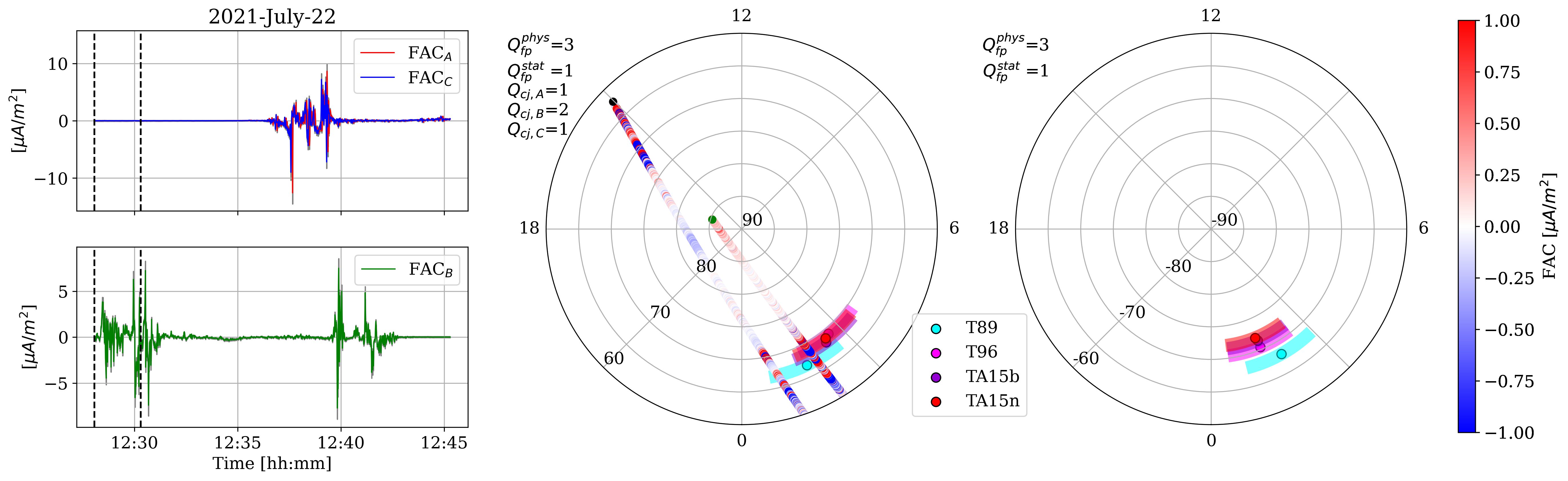}
    \caption{Example of BBF footpoint-Swarm conjunction for Swarm trace $\Delta t = 15$ minutes after end time of BBF detection. (Left) Swarm A, B and C FAC measurements, and the BBF duration indicated with the two vertical dashed lines. (Right) Polar plot with Swarm traces across the Northern and Southern Hemispheres and the corresponding FAC intensity. The BBF footpoint for the different Tsyganenko models are represented by circles, with shaded areas indicating the errors $Err_{MLAT} = \pm 1^\circ$, $Err_{MLT} = \pm 1$ h. The $Q_{fp}$ and $Q_{cj}$ indicate the footpoint quality and conjunction quality with Swarm A, B, and C.}
    \label{fig_swarm_bbf_example}
\end{figure}

We define a conjunction as occurring when the Swarm trace intersects the area defined by the BBF footpoint mapped to the ionosphere using any model, including their uncertainty margins. To evaluate the quality of these conjunctions, we introduce a conjunction quality metric ($Q_{cj}$). If the Swarm trace intersects two different Tsyganenko derived footpoints, the conjunction is assigned the highest quality ($Q_{cj}=1$); if it intersects only one of the footpoints, the conjunction quality is assigned $Q_{cj}=2$. Figure \ref{fig_swarm_bbf_example} shows an example of Swarm A, B and C traces during $\Delta t= 15~\mathrm{minutes}$, across the Southern and Northern Hemispheres, the corresponding FAC magnitude, and the BBF footpoint with the different Tsyganenko models. The shaded area of the footpoint is given by $Err_\mathrm{MLAT} = \pm 1^\circ$, and $Err_\mathrm{MLT} = \pm 1$~hour. In this example, we detect a conjunction in the Northern Hemisphere with Swarm A/C.

\begin{figure}
    \centering
    \includegraphics[width=\linewidth]{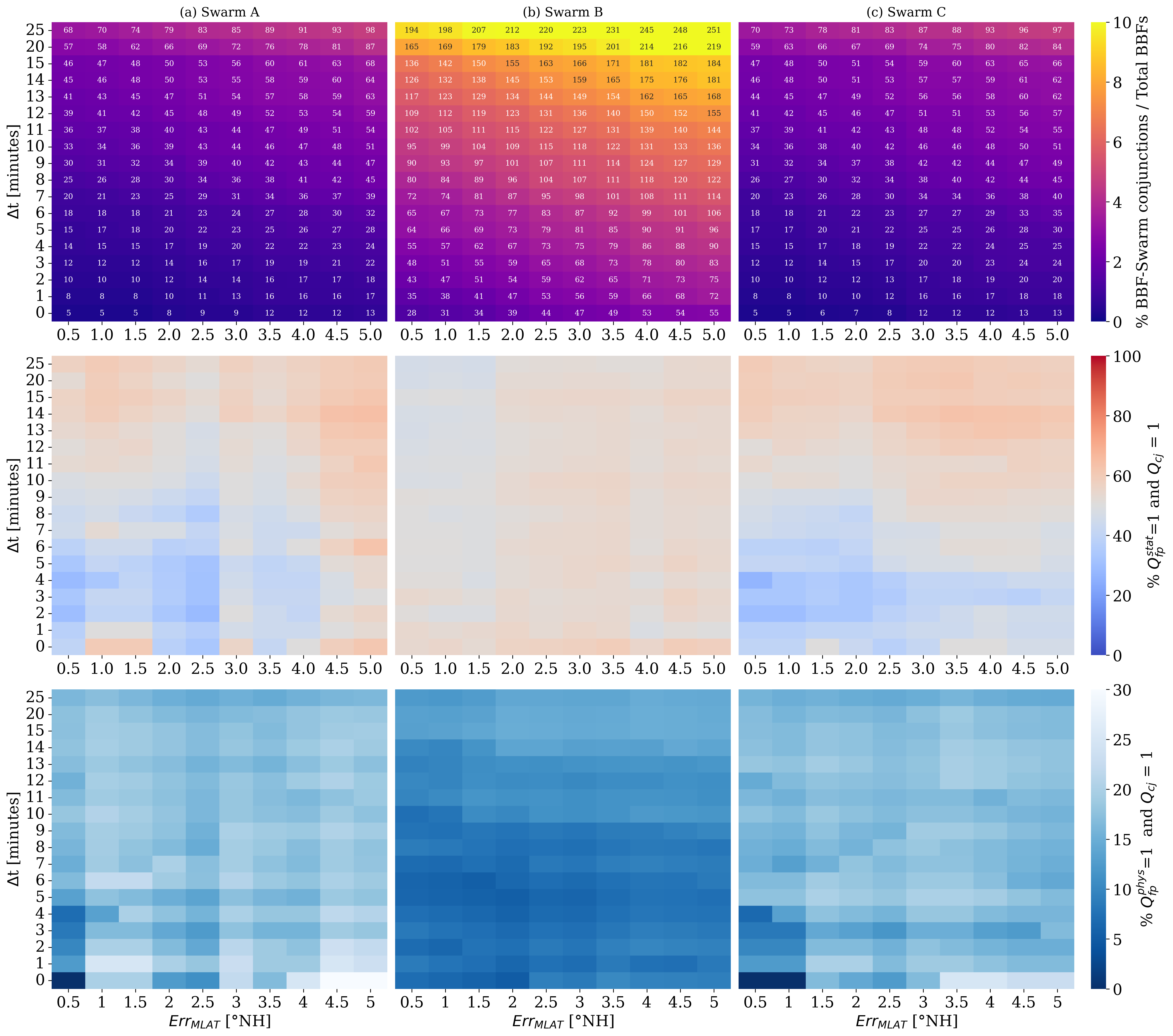}
    \caption{First row: Number of BBF footpoint-Swarm conjunctions for different time and MLAT position error ($\Delta t$, $\pm Err_\mathrm{MLAT}$ respectively) for the Northern Hemisphere and each Swarm satellite. The number inside each bin is the total number of conjunctions and the color is the percentage with respect to the total number of BBFs. Second row: Percentage of conjunctions with $Q_{fp}^{\rm{stat}}=1$ (the footpoint distance computed with the T89 and TA15n models is lower $\pm1$ hour in MLT and $\pm4^\circ$ in MLAT) and $Q_{cj}=1$ (Swarm satellite passing through two different footpoints). Third row: Similar to second row but with $Q_{fp}^{\rm{phys}}=1$ (the footpoint distance computed with the T89 and TA15n models is lower $\pm0.3$ hour in MLT and $\pm0.81^\circ$ in MLAT). Column (a) BBF conjunctions with Swarm A. Column (b) BBF conjunctions with Swarm B. Column (c) BBF conjunctions with Swarm C.}
    \label{fig_conj_heatmap}
\end{figure}

Figure \ref{fig_conj_heatmap} displays the number of Swarm-BBF footpoint conjunctions in the Northern Hemisphere and the fraction of $Q_1$ (i.e., $Q_{fp}=1$ and $Q_{cj}=1$, indicating that the footpoints computed by the T89 and TA15n models are in close alignment, with the Swarm satellite passing through two different footpoints) for two different criteria. The range for $\Delta t$ is estimated from the BBF database by assuming that the FACs are carried by Alfv{\'e}n waves \citep{Keiling_2009b}. The median length of the magnetic field line from the BBF position to the northward ionospheric footpoint is 25 $R_E$, and the median Alfv{\'e}n velocity of the BBFs is $470~\mathrm{km~s^{-1}}$. Assuming that the velocity is constant, the median time delay is $\Delta t=6~\mathrm{min}$. When considering the 10th and 90th percentiles of the field line length and Alfv{\'e}n velocity the extreme $\Delta t$ values are 1.9 and 24.5 minutes. The latitude range was selected up to $\pm 5^\circ$, which corresponds to a distance comparable to 1 hour MLT.

The results reveal that Swarm B shows a greater number of conjunctions compared to Swarm A and C. As expected, the number of conjunctions increases as the MLAT and time interval increases. In general, extending the time window by 1 minute results in a higher number of conjunctions than adjusting the MLAT error by $\pm 0.5^\circ$. From the two lower panels, we see that the fraction of $Q=1$ conjunctions with Swarm B is more homogeneous than Swarm A and C, which present a clear region of low fraction of $Q=1$ around $\Delta t = [0,6]$ minutes. When considering the size of the BBF footprint, we notice that in all cases less than half of the total number of conjunctions present $Q=1$. This highlights the challenges in achieving precise spatial alignment between BBF footpoints and satellite traces. Depending on the purpose of the research, different approaches may be appropriate for selecting conjunctions. If one is interested in detailed case studies, it is possible to work with very close conjunctions. However, if the goal is to conduct a statistical study, increasing the allowed MLAT error and extending the time window may be necessary to capture a broader range of conjunctions.

\section{Discussion}

Utilizing a database of 2394 busty bulk flows detected by MMS and applying six different Tsyganenko models, we mapped the BBFs to the Earth's ionosphere. On average, BBF footpoints are located in MLT coordinates around midnight, with a range of $\pm 3~\mathrm{hours}$, peaking in the pre-midnight sector (Figure \ref{fig_hist_foorprint}). This location aligns with regions of intense geomagnetic activity, likely associated with the substorm activity \cite{Weigel_2002}. 

The magnetic latitude of BBF footpoints are generally concentrated around $70^\circ \pm 5^\circ$ in both hemispheres. The Tsyganenko model 89, on average, returns a footpoint at lower latitudes compared with earlier models. This difference could be attributed to a more stretched magnetotail in the T89 model, as noted by \cite{Peredo_1993}, who reported that T89 is known to excessively stretch the magnetotail in the region $-10,R_E < X_{GSM} < -22,R_E$.

The MLT position of the footpoints exhibits a linear relationship with the $Y_\mathrm{GSM}$ position of the BBF, intersection MLT=0 at $Y_\mathrm{GSM}=0$. This means that on average the field lines represented by the Tsyganenko models do not cross the midnight meridian. Additionally, all Tsyganenko models that use $B_y$ and $B_z$ as input parameters (T96, T01, T04, T15b, and T15n) show a correlation with the IMF clock angle. For instance, Figure \ref{fig_mlt_ygsm} shows in the Northern Hemisphere that the linear regression shifts toward later MLT when $\theta_c\approx270^\circ$ and earlier for $\theta_c\approx90^\circ$. The opposite behavior is observed in the Southern Hemisphere.
 
The MLAT position is influenced by both the distance to Earth and the input parameter representing magnetospheric dynamics. As expected, more distant BBFs map, on average, to higher latitudes. Additionally, higher values of the magnetospheric dynamics parameter correspond to lower latitudes for the footpoints. This pattern aligns with the equatorward shift of the low-latitude boundary of the auroral oval during the growth phase of a substorm \cite{Akasofu_2017}. However, the parameters used to model magnetospheric dynamics differ across the various Tsyganenko models, leading to discrepancies in the latitude of the footpoints when one parameter indicates higher magnetospheric activity than another. Interestingly, T96 shows a low correlation with the Sym-h index ($r_p = -0.18$ and $r_s = -0.08$). This is likely because, during BBF events, the Sym-h index typically reflects non-storm periods (-30 nT $<$ Sym-h $<$ 0 nT). Similarly, the TA15 models exhibit negligible correlation between magnetic latitude and either the B-index or the N-index.

The relation of BBF-footpoints against the dipole tilt angle could not be explored in this work with the actual database. This should be taken into account when analyzing ionosphere and ground signatures. According to \cite{Eggington_2020}, the tilt angle has a strong effect on the location and intensity of magnetic reconnection which map down to the ionosphere. For instance, the polar cap contracts as the tilt angle increases, and FACs migrate to higher latitudes, exhibiting changes in morphology. To reduce the detection bias introduced by orbital effects, incorporating BBFs detected by other spacecraft, such as THEMIS and CLUSTER, can provide a more comprehensive and unbiased analysis \citep{Runov_2009, Cao_2006}.

\section{Conclusion}

In this paper, we provided an analysis of the variability in BBFs footpoint using six different Tsyganenko models and its impact on the number of BBF-Swarm conjunctions, with the following results:

\begin{itemize}
    \item Based on several widely used Tsyganenko magnetic field models, 90\% of the ionospheric footpoints of bursty bulk flow events inferred from MMS observations in the magnetotail are concentrated around $70^\circ \pm5^\circ$ MLAT and 0 MLT $\pm 3$ hours, with a peak in the pre-midnight sector.
    
    \item As expected, the BBF footpoint position in MLAT coordinate is closer to the pole as the BBF distance from Earth increases and closer to the equator for larger magnitudes of the Tsyganenko input parameter that quantifies the magnetospheric current systems (i.e, Kp, Dst, $E_{kv}$, $W_1$ for T89, T96, T01 and T04 respectively). However, the TA15 models do not exhibit this trend.
    
    \item The difference in MLAT between models shows footpoints at lower latitudes for the T89 model. This is related to the excessively stretched magnetotail in model T89.
    
    \item Footpoints of BBFs located at similar geocentric distances are found at lower MLAT for the model with the higher relative magnitude of the input parameter (Kp, Dst, $E_{kv}$, $W_1$ for the T89, T96, T01, and T04 models respectively) when both are normalized.

    \item The difference between the footpoints computed with the T89 and TA15n models is within the range of MLAT $\pm 4^\circ$, based on the 90th percentile of the data.
    
    \item The position of the BBF footpoint in the MLT coordinate is strongly correlated with the $Y_\mathrm{GSM}$ position of the BBF and the interplanetary magnetic field clock angle.
    
    \item The MLT differences between the models are within the range of $\pm 1$~hour, based on the 90th percentile of the data. The Northern Hemisphere footpoints of the T89 model in a similar $Y_\mathrm{GSM}$ position are located at later MLTs compared to the TA15n footpoint when $\theta_c\approx270^\circ$. In contrast, they appear at earlier MLT when $\theta_c\approx90^\circ$. An opposite trend is observed when comparing (T04 and TA15n) with TA15b. For the Southern Hemisphere, these relationships are reversed.

    \item Assuming that BBF-driven disturbances propagate along magnetic field lines at constant Alfv{\'e}n velocity, and using the median length of the field line to the northward ionospheric footpoint $L = 25R_E$, the median traveling time is $\Delta t=6~\mathrm{min}$. When considering the 10th and 90th percentiles of the field line length and Alfv{\'e}n velocity the extreme $\Delta t$ values are 1.9 and 24.5 minutes. 
        
    \item The number of BBF-Swarm conjunctions is larger with Swarm B than Swarm A/C, involving approximately 10\% of the total database. Moreover, Swarm B presents, on average, more conjunctions where the footpoints distance is lower than ($\pm 1$ hours in MLT and $\pm 4^\circ$ in MLAT) and Swarm path goes through both footpoints. However, the number of conjunctions is significantly reduced when considering a tighter footprint distance on the order of the BBF footpoint itself ($\pm 0.3$ hours in MLT and $\pm 0.81^\circ$ in MLAT).

\end{itemize}

\section{Acknowledgements}
VL, AD, LR, SB and OM would like to acknowledge the discussions with members of ISSI Team \#24-614 “Understanding ground magnetic disturbances due to field-aligned currents driven by magnetotail activity”. We thank the editor and two anonymous reviewer for evaluating this paper. We acknowledge use of NASA/GSFC's Space Physics Data Facility's OMNIWeb service, and OMNI data.

\section{Funding}
This work was supported by the ESA 4D ionosphere initiative. ESA contract No. 383 4000143412/23/I-EB and the Swedish Research Council (Grant 2021-06259). AD. received support from the Swedish National Space Agency (Grant 2020-00111). OM acknowledges support also by project MAGICS, ESA PRODEX contract 4000127660. 

\section{Data availability}
Solar wind parameters are available at \url{https://omniweb.gsfc.nasa.gov}. Solar wind driving parameters \textit{G} and \textit{W} at \url{https://rbsp-ect.newmexicoconsortium.org/data_pub/QinDenton/}, parameters \textit{Bindex} and \textit{Nindex} at \url{https://geo.phys.spbu.ru/~tsyganenko/empirical-models/magnetic_field/ta15}. IDL Geopack DLM is available at \url{https://korthhaus.com/idl-software/idl-geopack-dlm/}. The Hp30 index is accessible through the GFZ website at \url{https://kp.gfz.de/en/hp30-hp60/data}. The Database of MMS busty bulk flows including their ionospheric footpoints computed using Tsyganenko models (T89, T96, T01, T04, TA15B, and TA15N) is available at \url{https://doi.org/10.5281/zenodo.13789047}.

\begin{appendix} 
\section{Application of criteria to magnetic field line mapping}
\label{appendix_B}

\begin{figure}
    \centering
    \includegraphics[width=\linewidth]{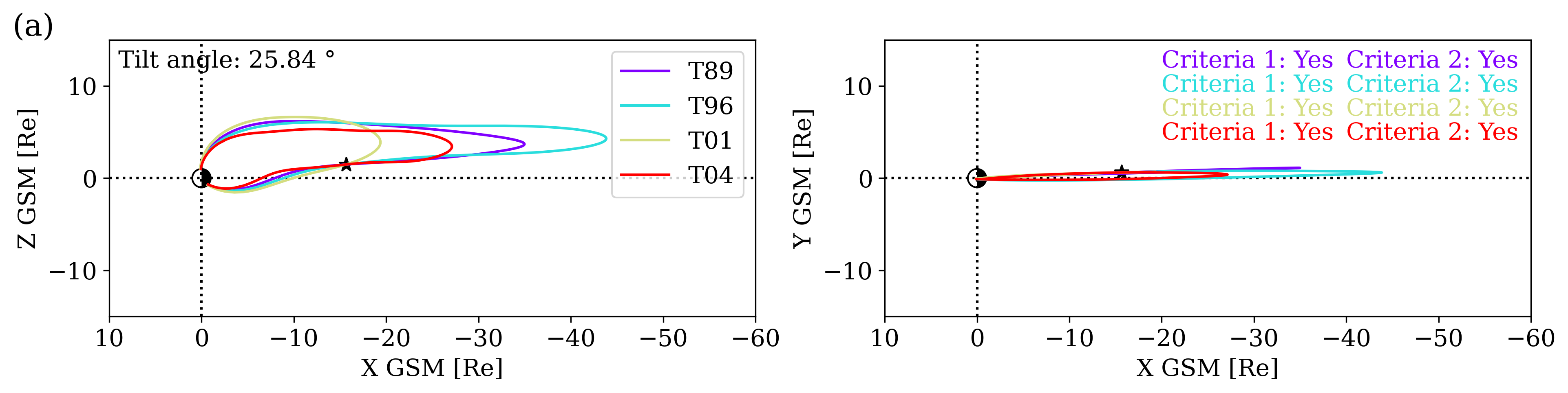}
    \includegraphics[width=\linewidth]{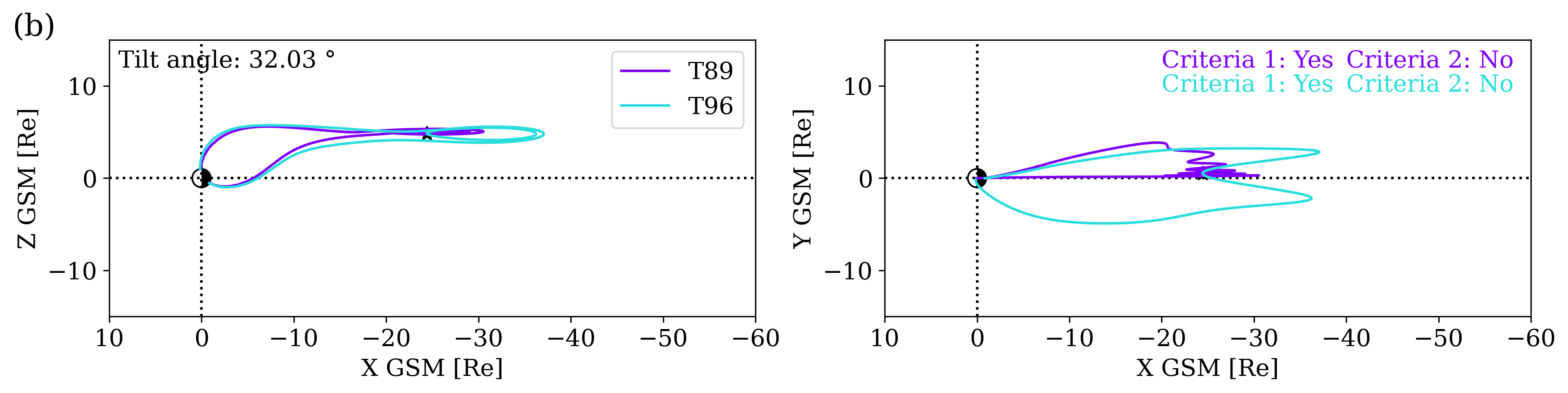}
    \includegraphics[width=\linewidth]{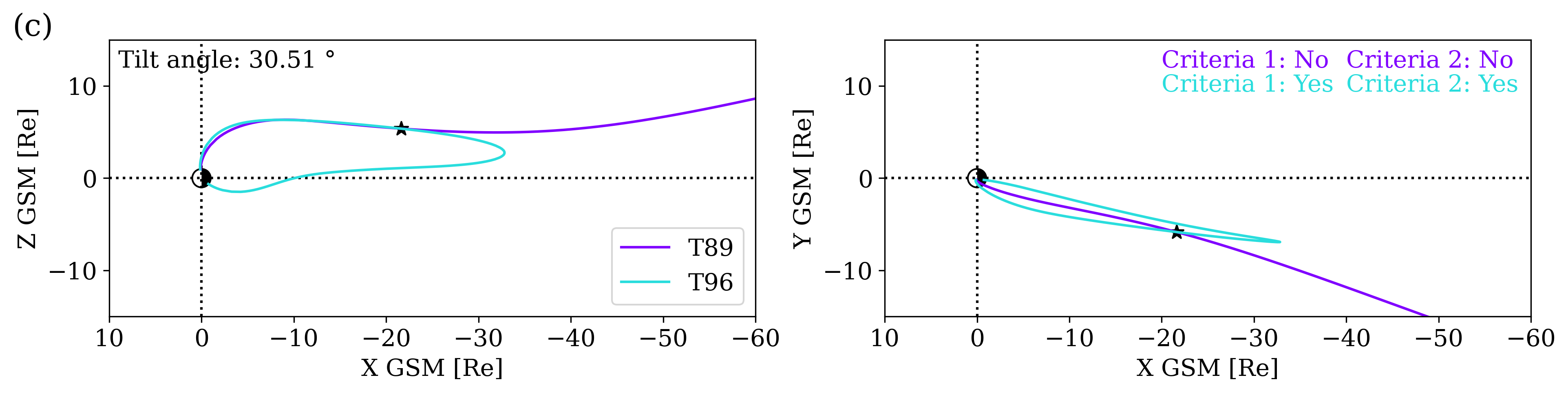}
    \caption{Example of three mapping scenarios when applying the Tsyganenko model to find the bursty bulk footpoint (indicated by an star). (a) The footpoint was calculated using four different Tsyganenko models, and all met both criteria. (b) Both Tsyganenko models (T89 and T96) returned both footpoints at the Earth's surface, but the shape of the field line did not met the second criterion (c) The footpoint computed with the T89 model did not return a footpoint in the Southern Hemisphere.}
    \label{fig_example_bad_model}
    
\end{figure}

In some cases, the footpoints obtained from the Tsyganenko model may be unreliable. Here, we present three different scenarios of field line tracing to illustrate this. In Figure \ref{fig_example_bad_model}a, four Tsyganenko models (i.e., T89, 96, T01 and T04) present both footpoints at the Earth's surface and the field line makes a single loop. Figure \ref{fig_example_bad_model}b presents both footpoints at the Earth's surface but perform several loops. Finally, \ref{fig_example_bad_model}c shows that the field line computed with T89 does not present a footpoint in the Southern Hemisphere.

\section{Estimate of the size of the footpoint}
\label{appendix_A}
To obtain a physical size of the footpoint to look for conjunctions, we consider that the BBFs correspond to an individual Dipolarizing Flux Bundle (DFB). From the conservation of magnetic flux along the DFB, we get $A_1 B_1 = A_2 B_2$, where $A_1$, $B_1$ and $A_2$, $B_2$ are the cross-sections of the DFB and magnetic field magnitudes at the BBF's location and Swarm's altitude, respectively. Assuming that the magnetic field at Swarm's altitude follows a dipole field model, we obtain 
\begin{equation}
    A_2 = \frac{B_1}{B_0 r^{-3} \sqrt{1 + 3\sin^2 \lambda}} A_1,
\end{equation}

where $B_0=3.15\times 10^4~\mathrm{T}$ the mean value of the magnetic field at the magnetic equator on the Earth's surface, $r=1 + h/R_E$ with $h=465~\mathrm{km}$ the altitude of the footpoint region, and $\lambda$ the magnetic latitude (MLAT). From MMS's observations of the BBFs, we get $B_1=12\pm 7~\mathrm{nT}$. From the mapping statistics (Figure \ref{fig_hist_foorprint}), we have $\lambda = 70^\circ \pm 4^\circ$. Finally, assuming a circular cross section with radius $R_1\approx 2~R_E$ \citep{Liu_2015}, we obtain $R_2\approx 110\pm 30~\mathrm{km}$.

\end{appendix}


\begin{thebibliography}{34}
\providecommand{\natexlab}[1]{#1}
\providecommand{\url}[1]{\texttt{#1}}
\providecommand{\urlprefix}{URL }
\providecommand{\eprint}[2][]{\url{#2}}

\bibitem[{{Akasofu}(2017)}]{Akasofu_2017}
{Akasofu}, S.-I., 2017.
\newblock {Auroral Substorms: Search for Processes Causing the Expansion Phase
  in Terms of the Electric Current Approach}.
\newblock \emph{Space Sci Rev}, \textbf{212}(1-2), 341--381.
\newblock Https://doi.org/10.1007/s11214-017-0363-7.

\bibitem[{{Angelopoulos} et~al.(1994){Angelopoulos}, {Kennel}, {Coroniti},
  {Pellat}, {Kivelson}, {Walker}, {Russell}, {Baumjohann}, {Feldman}, and
  {Gosling}}]{Angelopoulos_1994}
{Angelopoulos}, V., C.~F. {Kennel}, F.~V. {Coroniti}, R.~{Pellat}, M.~G.
  {Kivelson}, R.~J. {Walker}, C.~T. {Russell}, W.~{Baumjohann}, W.~C.
  {Feldman}, and J.~T. {Gosling}, 1994.
\newblock {Statistical characteristics of bursty bulk flow events}.
\newblock \emph{Journal of Geophysical Research}, \textbf{99}(A11),
  21,257--21,280.
\newblock Https://doi.org/10.1029/94JA01263.

\bibitem[{Aryan et~al.(2022)Aryan, Bortnik, Li, Weygand, Chu, and
  Angelopoulos}]{Aryan_2022}
Aryan, H., J.~Bortnik, J.~Li, J.~M. Weygand, X.~Chu, and V.~Angelopoulos, 2022.
\newblock Multiple conjugate observations of magnetospheric fast flow bursts
  using THEMIS observations.
\newblock \emph{Annales Geophysicae}, \textbf{40}(4), 531--544.
\newblock Https://doi.org/10.5194/angeo-40-531-2022.

\bibitem[{Boynton et~al.(2011)Boynton, Balikhin, Billings, Wei, and
  Ganushkina}]{Boynton_2011}
Boynton, R.~J., M.~A. Balikhin, S.~A. Billings, H.~L. Wei, and N.~Ganushkina,
  2011.
\newblock Using the NARMAX OLS-ERR algorithm to obtain the most influential
  coupling functions that affect the evolution of the magnetosphere.
\newblock \emph{Journal of Geophysical Research: Space Physics},
  \textbf{116}(A5).
\newblock Https://doi.org/10.1029/2010JA015505.

\bibitem[{{Burch} et~al.(2016){Burch}, {Moore}, {Torbert}, and
  {Giles}}]{Burch_2016}
{Burch}, J.~L., T.~E. {Moore}, R.~B. {Torbert}, and B.~L. {Giles}, 2016.
\newblock {Magnetospheric Multiscale Overview and Science Objectives}.
\newblock \emph{Space Science Reviews}, \textbf{199}(1-4), 5--21.
\newblock Https://doi.org/10.1007/s11214-015-0164-9.

\bibitem[{Cao et~al.(2006)Cao, Ma, Parks, Reme, Dandouras et~al.}]{Cao_2006}
Cao, J.~B., Y.~D. Ma, G.~Parks, H.~Reme, I.~Dandouras, et~al., 2006.
\newblock Joint observations by Cluster satellites of bursty bulk flows in the
  magnetotail.
\newblock \emph{Journal of Geophysical Research: Space Physics},
  \textbf{111}(A4).
\newblock Https://doi.org/10.1029/2005JA011322.

\bibitem[{Eggington et~al.(2020)Eggington, Eastwood, Mejnertsen, Desai, and
  Chittenden}]{Eggington_2020}
Eggington, J. W.~B., J.~P. Eastwood, L.~Mejnertsen, R.~T. Desai, and J.~P.
  Chittenden, 2020.
\newblock Dipole Tilt Effect on Magnetopause Reconnection and the Steady-State
  Magnetosphere-Ionosphere System: Global MHD Simulations.
\newblock \emph{Journal of Geophysical Research: Space Physics},
  \textbf{125}(7), e2019JA027,510.
\newblock Https://doi.org/10.1029/2019JA027510.

\bibitem[{Engebretson et~al.(2024)Engebretson, Gaffaney, Ochoa, Runov, Weygand
  et~al.}]{Engebretson_2024}
Engebretson, M.~J., S.~A. Gaffaney, J.~A. Ochoa, A.~Runov, J.~M. Weygand,
  et~al., 2024.
\newblock Signatures of Dipolarizing Flux Bundles in the Nightside Auroral
  Zone.
\newblock \emph{Journal of Geophysical Research: Space Physics},
  \textbf{129}(4), e2023JA032,266.
\newblock Https://doi.org/10.1029/2023JA032266.

\bibitem[{Juusola et~al.(2009)Juusola, Nakamura, Amm, and
  Kauristie}]{Juusola_2009}
Juusola, L., R.~Nakamura, O.~Amm, and K.~Kauristie, 2009.
\newblock Conjugate ionospheric equivalent currents during bursty bulk flows.
\newblock \emph{Journal of Geophysical Research: Space Physics},
  \textbf{114}(A4).
\newblock Https://doi.org/10.1029/2008JA013908.

\bibitem[{Keiling(2009)}]{Keiling_2009b}
Keiling, A., 2009.
\newblock Alfv{\'e}n waves and their roles in the dynamics of the Earth’s
  magnetotail: a review.
\newblock \emph{Space Science Reviews}, \textbf{142}, 73--156.

\bibitem[{Lanabere et~al.(2025)Lanabere, Richard, and Dimmock}]{Lanabere_2024}
Lanabere, V., L.~Richard, and A.~Dimmock, 2025.
\newblock {Database of MMS busty bulk flows including their ground and
  ionospheric footpoints}.
\newblock Dataset, 10.5281/zenodo.13789047,
  \urlprefix\url{https://doi.org/10.5281/zenodo.13789047}.

\bibitem[{Liu et~al.(2015)Liu, Angelopoulos, Chu, Zhou, and Yue}]{Liu_2015}
Liu, J., V.~Angelopoulos, X.~Chu, X.-Z. Zhou, and C.~Yue, 2015.
\newblock Substorm current wedge composition by wedgelets.
\newblock \emph{Geophysical Research Letters}, \textbf{42}(6), 1669--1676.
\newblock Https://doi.org/10.1002/2015GL063289.

\bibitem[{Liu et~al.(2013)Liu, Angelopoulos, Zhou, Runov, and Yao}]{Liu_2013b}
Liu, J., V.~Angelopoulos, X.-Z. Zhou, A.~Runov, and Z.~Yao, 2013.
\newblock On the role of pressure and flow perturbations around dipolarizing
  flux bundles.
\newblock \emph{Journal of Geophysical Research: Space Physics},
  \textbf{118}(11), 7104--7118.
\newblock Https://doi.org/10.1002/2013JA019256.

\bibitem[{Newell et~al.(2007)Newell, Sotirelis, Liou, Meng, and
  Rich}]{Newell_2007}
Newell, P.~T., T.~Sotirelis, K.~Liou, C.-I. Meng, and F.~J. Rich, 2007.
\newblock A nearly universal solar wind-magnetosphere coupling function
  inferred from 10 magnetospheric state variables.
\newblock \emph{Journal of Geophysical Research: Space Physics},
  \textbf{112}(A1).
\newblock Https://doi.org/10.1029/2006JA012015.

\bibitem[{Ngwira et~al.(2025)Ngwira, Nishimura, Weygand, Engebretson,
  Pulkkinnen, and Schuck}]{Ngwira_2025}
Ngwira, C.~M., Y.~Nishimura, J.~M. Weygand, M.~J. Engebretson, A.~Pulkkinnen,
  and P.~W. Schuck, 2025.
\newblock Observations of Localized Horizontal Geomagnetic Field Variations
  Associated With a Magnetospheric Fast Flow Burst During a Magnetotail
  Reconnection Event Detected by the THEMIS Spacecraft.
\newblock \emph{Journal of Geophysical Research: Space Physics},
  \textbf{130}(1), e2024JA032,651.
\newblock Https://doi.org/10.1029/2024JA032651.

\bibitem[{Opgenoorth et~al.(1994)Opgenoorth, Persson, Pulkkinen, and
  Pellinen}]{Opgenoorth_1994}
Opgenoorth, H.~J., M.~A. Persson, T.~I. Pulkkinen, and R.~J. Pellinen, 1994.
\newblock Recovery phase of magnetospheric substorms and its association with
  morning-sector aurora.
\newblock \emph{Journal of Geophysical Research}, \textbf{99}(A3).
\newblock Https://doi.org/10.1029/93JA01502.

\bibitem[{Peredo et~al.(1993)Peredo, Stern, and Tsyganenko}]{Peredo_1993}
Peredo, M., D.~P. Stern, and N.~A. Tsyganenko, 1993.
\newblock Are existing magnetospheric models excessively stretched?
\newblock \emph{Journal of Geophysical Research: Space Physics},
  \textbf{98}(A9), 15,343--15,354.
\newblock Https://doi.org/10.1029/93JA01150.

\bibitem[{Pulkkinen and Tsyganenko(1996)}]{Pulkkinen_1996}
Pulkkinen, T.~I., and N.~A. Tsyganenko, 1996.
\newblock Testing the accuracy of magnetospheric model field line mapping.
\newblock \emph{Journal of Geophysical Research: Space Physics},
  \textbf{101}(A12), 27,431--27,442.
\newblock Https://doi.org/10.1029/96JA02489.

\bibitem[{Qin et~al.(2007)Qin, Denton, Tsyganenko, and Wolf}]{Qin_2007}
Qin, Z., R.~E. Denton, N.~A. Tsyganenko, and S.~Wolf, 2007.
\newblock Solar wind parameters for magnetospheric magnetic field modeling.
\newblock \emph{Space Weather}, \textbf{5}(11).
\newblock Https://doi.org/10.1029/2006SW000296.

\bibitem[{Richard et~al.(2022{\natexlab{a}})Richard, Khotyaintsev, Graham, and
  Russell}]{Richard_2022}
Richard, L., Y.~V. Khotyaintsev, D.~B. Graham, and C.~T. Russell,
  2022{\natexlab{a}}.
\newblock Are Dipolarization Fronts a Typical Feature of Magnetotail Plasma
  Jets Fronts?
\newblock \emph{Geophysical Research Letters}, \textbf{49}(22), e2022GL101,693.
\newblock Https://doi.org/10.1029/2022GL101693.

\bibitem[{Richard et~al.(2022{\natexlab{b}})Richard, Khotyaintsev, Graham, and
  Russell}]{Richard_2022db}
Richard, L., Y.~V. Khotyaintsev, D.~B. Graham, and C.~T. Russell,
  2022{\natexlab{b}}.
\newblock {Are Dipolarization Fronts a Typical Feature of Magnetotail Plasma
  Jets Fronts? [Data set]}.
\newblock 10.5281/zenodo.7009706,
  \urlprefix\url{https://doi.org/10.5281/zenodo.7009706}.

\bibitem[{Runov et~al.(2009)Runov, Angelopoulos, Sitnov, Sergeev, Bonnell,
  McFadden, Larson, Glassmeier, and Auster}]{Runov_2009}
Runov, A., V.~Angelopoulos, M.~I. Sitnov, V.~A. Sergeev, J.~Bonnell, J.~P.
  McFadden, D.~Larson, K.-H. Glassmeier, and U.~Auster, 2009.
\newblock THEMIS observations of an earthward-propagating dipolarization front.
\newblock \emph{Geophysical Research Letters}, \textbf{36}(14).
\newblock Https://doi.org/10.1029/2009GL038980.

\bibitem[{Shevchenko et~al.(2010)Shevchenko, Sergeev, Kubyshkina, Angelopoulos,
  Glassmeier, and Singer}]{Shevchenko_2010}
Shevchenko, I.~G., V.~Sergeev, M.~Kubyshkina, V.~Angelopoulos, K.~H.
  Glassmeier, and H.~J. Singer, 2010.
\newblock Estimation of magnetosphere-ionosphere mapping accuracy using
  isotropy boundary and THEMIS observations.
\newblock \emph{Journal of Geophysical Research: Space Physics},
  \textbf{115}(A11).
\newblock Https://doi.org/10.1029/2010JA015354.

\bibitem[{Tsyganenko(1987)}]{Tsyganenko_1987}
Tsyganenko, N., 1987.
\newblock Global quantitative models of the geomagnetic field in the cislunar
  magnetosphere for different disturbance levels.
\newblock \emph{Planetary and Space Science}, \textbf{35}(11), 1347--1358.
\newblock Https://doi.org/10.1016/0032-0633(87)90046-8.

\bibitem[{{Tsyganenko}(1989)}]{Tsyganenko_1989}
{Tsyganenko}, N.~A., 1989.
\newblock {A magnetospheric magnetic field model with a warped tail current
  sheet}.
\newblock \emph{Planetary and Space Science}, \textbf{37}(1), 5--20.
\newblock Https://doi.org/10.1016/0032-0633(89)90066-4.

\bibitem[{{Tsyganenko}(1996)}]{Tsyganenko_1996}
{Tsyganenko}, N.~A., 1996.
\newblock {Effects of the solar wind conditions in the global magnetospheric
  configurations as deduced from data-based field models (Invited)}.
\newblock In E.~J. {Rolfe} and B.~{Kaldeich}, eds., International Conference on
  Substorms, vol. 389 of \emph{ESA Special Publication}, 181.

\bibitem[{{Tsyganenko}(2002{\natexlab{a}})}]{Tsyganenko_2002a}
{Tsyganenko}, N.~A., 2002{\natexlab{a}}.
\newblock {A model of the near magnetosphere with a dawn-dusk asymmetry 1.
  Mathematical structure}.
\newblock \emph{Journal of Geophysical Research (Space Physics)},
  \textbf{107}(A8), 1179.
\newblock Https://doi.org/10.1029/2001JA000219.

\bibitem[{{Tsyganenko}(2002{\natexlab{b}})}]{Tsyganenko_2002b}
{Tsyganenko}, N.~A., 2002{\natexlab{b}}.
\newblock {A model of the near magnetosphere with a dawn-dusk asymmetry 2.
  Parameterization and fitting to observations}.
\newblock \emph{Journal of Geophysical Research (Space Physics)},
  \textbf{107}(A8), 1176.
\newblock Https://doi.org/10.1029/2001JA000220.

\bibitem[{Tsyganenko and Andreeva(2015)}]{Tsyganenko_2015}
Tsyganenko, N.~A., and V.~A. Andreeva, 2015.
\newblock A forecasting model of the magnetosphere driven by an optimal solar
  wind coupling function.
\newblock \emph{Journal of Geophysical Research: Space Physics},
  \textbf{120}(10), 8401--8425.
\newblock Https://doi.org/10.1002/2015JA021641.

\bibitem[{{Tsyganenko} and {Sitnov}(2005)}]{Tsyganenko_2005}
{Tsyganenko}, N.~A., and M.~I. {Sitnov}, 2005.
\newblock {Modeling the dynamics of the inner magnetosphere during strong
  geomagnetic storms}.
\newblock \emph{Journal of Geophysical Research (Space Physics)},
  \textbf{110}(A3), A03208.
\newblock Https://doi.org/10.1029/2004JA010798.

\bibitem[{{Vasyliunas} et~al.(1982){Vasyliunas}, {Kan}, {Siscoe}, and
  {Akasofu}}]{Vasylundas_1982}
{Vasyliunas}, V.~M., J.~R. {Kan}, G.~L. {Siscoe}, and S.~I. {Akasofu}, 1982.
\newblock {Scaling relations governing magnetospheric energy transfer}.
\newblock \emph{Planetary and Space Science}, \textbf{30}(4), 359--365.
\newblock Https://doi.org/10.1016/0032-0633(82)90041-1.

\bibitem[{Wei et~al.(2021)Wei, Dunlop, Yang, Dong, Yu, and Wang}]{Wei_2021}
Wei, D., M.~W. Dunlop, J.~Yang, X.~Dong, Y.~Yu, and T.~Wang, 2021.
\newblock Intense dB/dt Variations Driven by Near-Earth Bursty Bulk Flows
  (BBFs): A Case Study.
\newblock \emph{Geophysical Research Letters}, \textbf{48}(4), e2020GL091,781.
\newblock Https://doi.org/10.1029/2020GL091781.

\bibitem[{Weigel et~al.(2002)Weigel, Vassiliadis, and Klimas}]{Weigel_2002}
Weigel, R.~S., D.~Vassiliadis, and A.~J. Klimas, 2002.
\newblock Coupling of the solar wind to temporal fluctuations in ground
  magnetic fields.
\newblock \emph{Geophysical Research Letters}, \textbf{29}(19), 21--1--21--4.
\newblock Https://doi.org/10.1029/2002GL014740.

\bibitem[{{Yamazaki} et~al.(2022){Yamazaki}, {Matzka}, {Stolle},
  {Kervalishvili}, {Rauberg}, {Bronkalla}, {Morschhauser}, {Bruinsma},
  {Shprits}, and {Jackson}}]{Yamazaki_2022}
{Yamazaki}, Y., J.~{Matzka}, C.~{Stolle}, G.~{Kervalishvili}, J.~{Rauberg},
  O.~{Bronkalla}, A.~{Morschhauser}, S.~{Bruinsma}, Y.~Y. {Shprits}, and D.~R.
  {Jackson}, 2022.
\newblock {Geomagnetic Activity Index Hpo}.
\newblock \emph{Geophysical Research Letters}, \textbf{49}(10), e98860.
\newblock Https://doi.org/10.1029/2022GL098860.

\end{thebibliography}


\end{document}